\documentclass[vecphys]{svmult}

\usepackage{makeidx}         
\usepackage{graphicx}        
\usepackage{multicol}        
\usepackage[bottom]{footmisc}

\makeindex             

\begin{document}

\title*{Environment of compact extragalactic radio sources}
\author{A. Labiano\inst{1},
C.P.O'Dea\inst{2},
P.D. Barthel\inst{3}\and
R.C. Vermeulen\inst{4}}
\institute{Departamento de Astrof\'isica Molecular e Infrarroja, Instituto de Estructura de la Materia (CSIC), Madrid, Spain
\texttt{labiano@damir.iem.csic.es}
\and Department of Physics, Rochester Institute of Technology, Rochester, NY, 14623, USA \texttt{odea@cis.rit.edu}
\and Kapteyn Astronomical Institute, Groningen, 9700 AV, The Netherlands \texttt{pdb@astro.rug.nl}
\and Netherlands Foundation for Research in Astronomy (ASTRON), PO Box 2, 7990 AA Dwingeloo, The Netherlands \texttt{rvermeulen@astron.nl}}

%
%
\maketitle
\index{Author1}
\index{Author2}
\begin{abstract}
We have studied the interrelation of young AGN with their hosts. The objects of study are the young and powerful GPS and CSS radio sources. Due to their small size, GPS and CSS sources are excellent probes of this relation. Furhthermore, their young age allows us to compare them to the larger, old radio sources and establish a time-line evolution of this relation. Combining imaging and spectroscopy at UV, optical and radio wavelengths we find evidence of strong interaction between the host and the radio source. The presence and expansion of the radio source clearly affects the properties and evolution of the host. Furthermore, the radio source and host significantly affect each other's evolution.  We describe our results and how these interactions take place.
\end{abstract}

\section{Introduction}
\label{sec:intro}



 GHz Peaked Spectrum (GPS) and Compact Steep Spectrum (CSS) radio sources are the most likely candidates for the progenitors of the large scale powerful classical double (FR) sources (e.g., \cite{O'Dea91,Fanti90,Fanti95,Readhead96a,Readhead96b,O'Dea97}; for a review see \cite{O'Dea98}). The GPS and CSS sources are powerful but compact radio sources whose spectra are generally simple and convex with peaks near 1 GHz and 100 MHz respectively. The GPS sources are contained within the  extent of the optical narrow emission line region ($< 1$ kpc) while the CSS sources are contained within the host galaxy ($< 15$ kpc).\\

Current models for the evolution of powerful radio galaxies suggest that these sources propagate from the $\sim 10$ pc to Mpc scales at roughly constant velocity through an ambient medium which declines in density as $\rho(R) \propto R^{-2}$ while the sources decline in radio luminosity as $L_{rad} \propto R^{-0.5}$ (\cite{Fanti95,Begelman96,Readhead96b,Young97,Kaiser97a,Kaiser97b,Snellen00}). Such a scenario is consistent with the observed number densities of powerful radio sources as a function of linear size (from tens of parsecs to hundreds of kpc; e.g., \cite{O'Dea97,Fanti01}). However, the situation must be more complicated than this simple picture. (1) The GPS and CSS sources must interact with the host galaxy as they propagate through it. The discovery of emission line gas aligned with and presumably co-spatial with the CSS radio sources and the presence of broad and complex integrated emission line profiles (\cite{Gelderman94}) indicates that the radio source is strongly interacting with the ambient gas (\cite{Vries97,Vries99,Axon00}). Therefore, we would expect shocks to contribute strongly to the ionization of the gas (\cite{Bicknell97}). (2) The GPS sources are observed to have expansion velocities several times higher than the estimated advance speeds of large scale classical doubles (\cite{Alexander87}). This would require the evolving GPS sources to decelerate as they propagate though the host galaxy and would require the radio sources to dim faster than the simple models predict. It may be that the deceleration takes place via interaction with ambient gas (see \cite{Young93,Carvalho94,Carvalho98}). We have carried out a multiwavelength study of how GPS and CSS sources evolve and relate with their host as the expansion takes place.

\section{Emission line gas}
\label{sec:elg}

The first traces of interaction between GPS/CSS and their hosts were find in the emission line gas. The properties observed in [O III] $\lambda$5007 emission lines (\cite{Gelderman94}) suggested that, even though the AGN was partly photoionizing the nebula, the radio source was dominating the emission-line kinematics. A few years later, the alignment effect (similar extent for radio and optical wavelengths) seen in large radio sources (e.g., \cite{Chambers87, McCarthy87}) was found in GPS and CSS (\cite{Vries97, Vries99, Axon00}). The alignment between the radio source and emission line gas suggests interaction between them, as the radio source propagates through the host. The emission line gas was found to be brighter in the center of the radio lobes and dimmed with distance. Therefore, the hypothesis adopted was that the gas inside the radio lobe had been shock-ionized while the gas outside was photoionized by the precursor gas in the shock. At the same time, the AGN could contribute to the whole system with photoionization. Cooling time arguments suggested lobe expansion velocities $>$1000 km s$^{-1}$ \, for most of the sources. \cite{O'Dea02} studied the kinematics of the emission line nebulae of CSS sources, finding that the radio source is accelerating the clouds of gas in the host. \\

We have obtained Hubble Space Telescope Imaging Spectrograph long-slit spectroscopy of the aligned emission line nebulae in three compact steep spectrum radio sources: 3C~67, 3C~277.1, and 3C~303.1. Comparing our data with with shock and photo-ionization models (MAPPINGS \cite{Kewley03,Dopita96} and CLOUDY \cite{Ferland98}) and examining the relationships between the ionization diagnostics and the gas kinematics, we find that:\\


1.- In general the extended emission lines are consistent with a mixture of shocked and photo-ionized gas. In 3C~67, the data lie between the MAPPINGS models for a contribution to the observed luminosity from shocked gas ranging from 0 to 50$\%$. In 3C~303.1, the contribution to the luminosity from shocked gas is between 30 and 70$\%$. In 3C~277.1, the data scatter around the model for 100$\%$ contribution to the luminosity from precursor gas. The sources tend to lie in the regions for moderate to high shock velocities (500 to 1000 km/s) and we do not obtain any useful constraints on magnetic field strengths in the MAPPINGS models.\\

2.- The three sources show a decrease in ionization with distance from the nucleus (consistent with a decrease in photoionization with distance) a weak trend for ionization to decrease with increasing FWHM and a strong trend for ionization to decrease with increasing velocity offset (which is consistent with shock ionization). \\

These results are consistent with a picture in which the CSS sources interact with dense clouds as they propagate through their host galaxies, shocking the clouds thereby ionizing and accelerating them (as suggested by \cite{O'Dea02}).

\section{Cold gas}
\label{sec:cg}

Many GPS, CSS and larger radio sources are in merging or interacting systems (e.g., \cite{Vries00, Johnston05}), therefore we expect their hosts to have dense nuclear environments. Furthermore, most of the radio loud AGN seem to live in elliptical galaxies (e.g., \cite{Dunlop03}). \cite{Walsh89} found that the presence of gas (and dust) in early type galaxies is correlated with the occurrence and strength of a central radio source. This gas is found both in molecular and atomic form (e.g., \cite{Knapp96, Oosterloo99}). Observations at different wavelengths find central structures of gas in AGN hosts (\cite{Gorkom89, Verdoes99, Evans99, Langevelde00, Morganti01}). If the gas is centrally concentrated and GPS/CSS sources are precursors of FR sources, we expect GPS/CSS to show higher densities in the gas surrounding their radio lobes\footnote{However, extremely high densities may confine the sources.}. In 2003, using the recently improved Westerbork Synthesis Radio Telescopes, \cite{Vermeulen03} and \cite{Pihlstrom03} presented a study of the occurrence and properties of atomic gas associated with compact radio sources. They found that $\sim50\%$ of GPS and CSS sources showed HI 21 cm absorption, in contrast to normal elliptical galaxies ($< 10\%$, \cite{Gorkom89}). Furthermore, they found that GPS sources tend to have higher HI column densities than CSS sources and that these densities were consistent with the young scenario. However, their data lack spatial resolution to accurately locate the HI absorption. Just a few high resolution HI observations of GPS/CSS sources were available at that time (\cite{Conway96, Peck98, Peck99, Peck02}). \\

We have used European VLBI Network UHF band spectral line observations, to localise the redshifted 21cm HI absorption known to occur in the subgalactic sized compact steep spectrum galaxies 3C~49 and 3C~268.3. The radio lobes with detected HI absorption (1) are brighter and closer to the core than the opposite lobes; (2) are more depolarized; and (3) are preferentially associated with optical emission line gas. The association between the HI absorption and the emission line gas, supports the hypothesis that the HI absorption is produced in the atomic cores of the emission line clouds. However, we cannot rule out the existence of HI elsewhere. We suggest that the asymmetries in the radio and optical emission are due to interaction of the radio source with an asymmetric distribution of dense clouds in their environment. Our results are consistent with a picture in which CSS sources interact with clouds of dense gas as they propagate through their host galaxy.\\



\section{Star formation and AGN - Starburst connection}
\label{sec:sf}

Mergers and strong interactions can trigger AGN activity in a galaxy (e.g., \cite{Heckman86, Baum92, Israel98}). These events can also produce instabilities in the ISM and trigger star formation (e.g., \cite{Ho05}). Numerical simulations and models (e.g., \cite{Mellema02,Rees89}) suggest that the advancement of the jets through the host galaxy environment can also trigger star formation. Imaging studies in ultraviolet (UV) light of the hosts of large 3CR sources find traces of episodes of star formation around the time when the radio source was triggered (\cite{Koekemoer99, Allen02, Chiaberge02, O'Dea01, O'Dea03, Martel02}) suggesting a possible link between both. \\

GPS and CSS radio sources are young, smaller versions of the large powerful radio sources, so they are expected to exhibit signs of more recent star formation.  They have not yet completely broken through the host ISM, so these interactions are expected to be even more important than in the larger sources. Near-UV observations are very sensitive to the presence of hot young stars and therefore will trace recent star formation events. We have obtained high resolution HST - Advanced Camera for Surveys near-UV images of GPS and CSS sources to study the morphology and the extent of recent star formation.\\

We find near UV emission consistent with the presence of recent star formation in most of the observed the sources. In the CSS sources 1443+77 and 1814--637 the near UV light is aligned with and is co-spatial with the radio source. We suggest that in these sources the UV light is produced by star formation triggered and/or enhanced by the radio source. \\

The connection between the AGN and the star formation is not yet clear but our observations suggest that some connection exists. Stellar synthesis models are consistent with a burst of star formation before the formation of the radio source. The starburst and AGN activity could have been triggered by the same event. However, observations at other wavelengths and measurement of the colors are needed to definitively establish the nature of the observed UV light.

\section{Conclusions}
\label{sec:con}

We have used the young and powerful GPS and CSS radio sources to study the interrelation of AGN with their hosts. We find evidence of strong interaction between the radio source and the host at all wavelengths studied, suggesting that the presence and expansion of powerful radio sources clearly affect the properties and evolution of their hosts.\\

The radio source and host can significantly affect each others evolution. However, this influence takes place in different ways. The influence that the host has on the radio source is somehow indirect, although it can completely change its destiny: depending of the contents, distribution and density of the gas, the radio source will die early, expand and grow into the large FR sources, or remain confined inside its host. In contrast, the influence of the radio source on its host seems to be more direct and takes place during its expansion through the host: the radio source will affect the kinematics and ionization of the emission line gas, and may change the star formation history of the host.\\


\def\etal{{\it et al.~}}

\newcommand\aj{AJ~}
\newcommand\apj{ApJ~}
\newcommand\apjl{ApJ~}
\newcommand\apjs{ApJS~}
\newcommand\aap{A\&A~}
\newcommand\aapr{A\&A~Rev.~}
\newcommand\aaps{A\&AS}
\newcommand\mnras{MNRAS~}
\newcommand\pasa{PASA~}
\newcommand\pasp{PASP~}
\newcommand\nat{Nature~}
\newcommand\nar{New A Rev.~}


\printindex
\end{document}